\documentclass[pra,floatfix,footinbib,superscriptaddress, twocolumn, amsmath,amssymb,amsfonts]{revtex4-1} 

\usepackage[T1]{fontenc}
\usepackage[latin9]{inputenc}
\usepackage{lmodern} 
\usepackage{color}
\usepackage{bbold}
\usepackage{dsfont}
\usepackage{graphicx}
\usepackage[caption=false]{subfig}
\usepackage[colorlinks]{hyperref}
\usepackage[bottom]{footmisc}
\usepackage{enumerate}
\usepackage{float}
\usepackage{mathtools}
\usepackage{leftindex}
\usepackage{empheq}
\usepackage{tikz}

\usepackage[normalem]{ulem}
\DeclareMathAlphabet\mbc{OMS}{cmsy}{b}{n}

\usepackage{balance}

\begin{document}

\global\long\def\eqn#1{\begin{align}#1\end{align}}
\global\long\def\vec#1{\overrightarrow{#1}}
\global\long\def\ket#1{\left|#1\right\rangle }
\global\long\def\bra#1{\left\langle #1\right|}
\global\long\def\bkt#1{\left(#1\right)}
\global\long\def\sbkt#1{\left[#1\right]}
\global\long\def\cbkt#1{\left\{#1\right\}}
\global\long\def\abs#1{\left\vert#1\right\vert}
\global\long\def\cev#1{\overleftarrow{#1}}
\global\long\def\der#1#2{\frac{{d}#1}{{d}#2}}
\global\long\def\pard#1#2{\frac{{\partial}#1}{{\partial}#2}}
\global\long\def\re{\mathrm{Re}}
\global\long\def\im{\mathrm{Im}}
\global\long\def\dd{\mathrm{d}}
\global\long\def\ddd{\mathcal{D}}
\global\long\def\hmb#1{\hat{\mathbf #1}}
\global\long\def\avg#1{\left\langle #1 \right\rangle}
\global\long\def\mr#1{\mathrm{#1}}
\global\long\def\mb#1{{\mathbf #1}}
\global\long\def\mc#1{\mathcal{#1}}
\global\long\def\tr{\mathrm{Tr}}
\global\long\def\dbar#1{\Bar{\Bar{#1}}}

\global\long\def\nth{$n^{\mathrm{th}}$\,}
\global\long\def\mth{$m^{\mathrm{th}}$\,}
\global\long\def\non{\nonumber}

\newcommand{\orange}[1]{{\color{orange} {#1}}}
\newcommand{\cyan}[1]{{\color{cyan} {#1}}}
\newcommand{\blue}[1]{{\color{blue} {#1}}}
\newcommand{\yellow}[1]{{\color{yellow} {#1}}}
\newcommand{\green}[1]{{\color{green} {#1}}}
\newcommand{\red}[1]{{\color{red} {#1}}}
\global\long\def\todo#1{\orange{{$\bigstar$ \cyan{\bf\sc #1}}$\bigstar$} }

\title{Spontaneous Emission in the presence of Quantum Mirrors}
\author{Kanu Sinha}
\email{kanu@arizona.edu}
\affiliation{Wyant College of Optical Sciences and Department of Physics, University of Arizona, Tucson, AZ 85719}
\author{Jennifer Parra-Contreras}
\affiliation{Wyant College of Optical Sciences and Department of Physics, University of Arizona, Tucson, AZ 85719}\author{Annyun Das}
\affiliation{Wyant College of Optical Sciences and Department of Physics, University of Arizona, Tucson, AZ 85719}
\author{Pablo Solano}
\email{psolano@udec.cl}
\affiliation{Departamento de F\'{i}sica, Facultad de Ciencias F\'{i}sicas y Matem\'{a}ticas, Universidad de Concepci\'{o}n, Concepci\'{o}n,
Chile}

\begin{abstract}
Arrays of atoms coupled to waveguides can behave as mirrors. We consider an array of $ \Lambda$-type three-level atoms wherein preparing the atoms in one ground state or another leads to reflection or transmission of the guided electromagnetic field; a superposition of the two ground states thus corresponds to a  coherent superposition of mirror-like and transparent boundary conditions. We analyze the spontaneous emission of an excited two-level atom in the presence of such a \textit{quantum mirror}, and inside a cavity formed by quantum mirrors, demonstrating that the resulting dynamics of the excited atom can exhibit exotic features, e.g., a superposition of Rabi cycle and exponential decay. Our results pave the way for exploring quantum electrodynamics (QED) phenomena in a paradigm wherein boundary conditions can exhibit quantum superpositions and correlations. 
 \end{abstract}

\maketitle

\textit{Introduction}.--- The modification of the quantized electromagnetic (EM) field in the presence of boundaries is a ubiquitous  aspect of Quantum Electrodynamics  (QED), and integral to a wide range of phenomena  from cavity QED~\cite{Walther2006, Mabuchi2002}, to nanophotonics~\cite{NovotnyBook,Damico2019}, and  Casimir physics~\cite{Milonni}. While boundary conditions  are typically described in terms of the macroscopic polarization induced in a medium in the presence of an applied EM field~\cite{Jackson}, their emergence  can be more rigorously established from a microscopic treatment of media as a continuous distribution of atomic scatterers~\cite{CookMilonni1987, Fearn1996, Feng1990}.

 The remarkable progress in experimental control of atomic systems has enabled creation of materials at an atom-by-atom scale~\cite{ChangRMP}.  For example, arrays of atoms arranged in a Bragg configuration can exhibit strong reflection on an incident EM field. Such atomic Bragg mirrors have been explored theoretically and experimentally  with  atoms trapped in free-space optical lattices~\cite{Deutsch1995,Weidemuller1995, Birkl1995,Schilke2011,Slama2005}, as well as near waveguides, enabling enhanced light-matter interactions, and thereby a strong and coherent reflection of light~\cite{Chang2011, LeKien2014}.  Such systems are relevant to implementations of various quantum information processing protocols~\cite{Chang2012,GonzalezTudela2015, Paulisch2016, AsenjoGarcia2017, GonzalezTudela2017, Shah24}.

As a striking phenomenon, recent experiments with subwavelength Rydberg atom arrays and superconducting circuits have demonstrated coherent switching of the optical response of  boundary conditions realized by atomic, or atom-like, mirrors \cite{Srakaew2023, Mirhosseini2019}. These experiments bring forth the novel prospect of realizing coherent quantum superpositions of two disparate  boundary conditions on the EM field. The emergent boundary conditions in the presence of such atomic arrays must account for the fact that the macroscopic polarization induced in the medium can exist in a quantum superposition. The consequent QED phenomena in the presence of such ``quantum boundary conditions'' are, as yet, largely unexplored.

 In this letter we propose and analyze a system with \textit{quantum  mirrors},  wherein the emergent optical properties of an atomic array  depend on the specific quantum states of its constituent atoms.  Preparing the array of atoms in a macroscopic quantum superposition of the two states that correspond to distinct boundary conditions on the EM field, we analyze the spontaneous emission of a two-level atom near such a quantum mirror (QM), and inside a cavity formed by two QMs. 
 \begin{figure}[b]
     \centering
     \includegraphics[width = 0.48\textwidth]{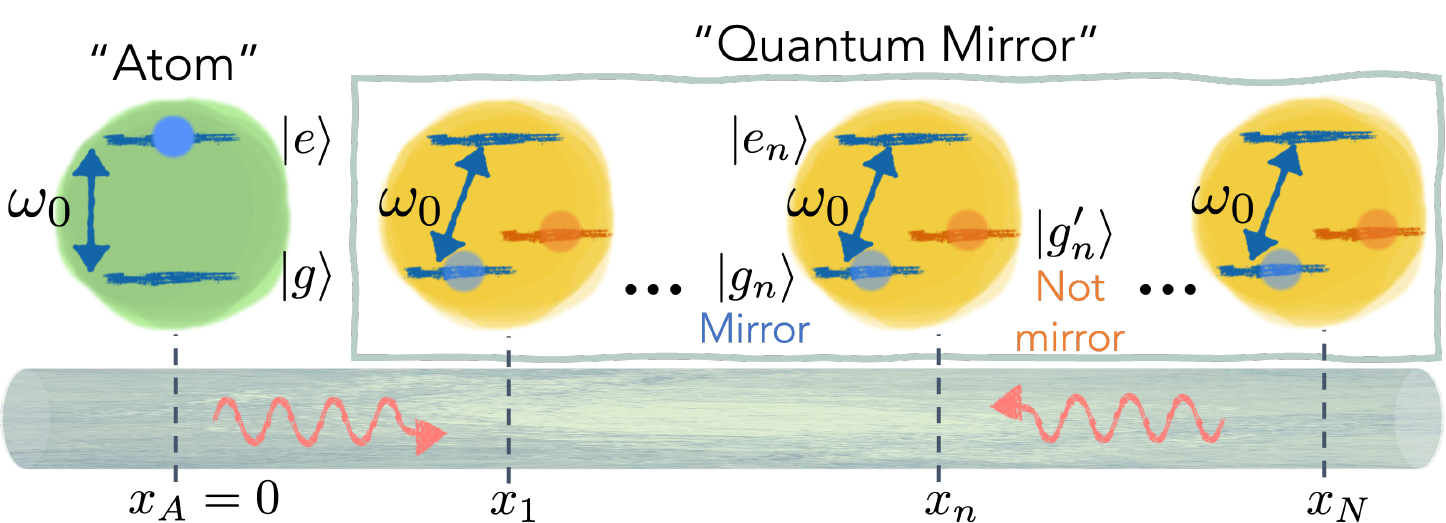}
     \caption{Schematic representation of a two-level atom (``Atom'') in front of an atomic array of $N$ $\Lambda$-type three-level atoms (``Quantum Mirror'' (QM))  coupled via a waveguide. The atomic array is arranged in a Bragg configuration with $|x_n - x_{n-1}| = \lambda_0/2$, such that for all atoms in the state $ \ket{g_n}$, the array behaves as a mirror. The level $ \ket{g'_n}$ being uncoupled from other atomic levels,  the array  with all atoms in $ \ket{g_n'}$ acts transparently.}
     \label{Fig:SingleQM1}
 \end{figure}

\begin{figure*}[t]
    \centering
    \includegraphics[width =  \textwidth]{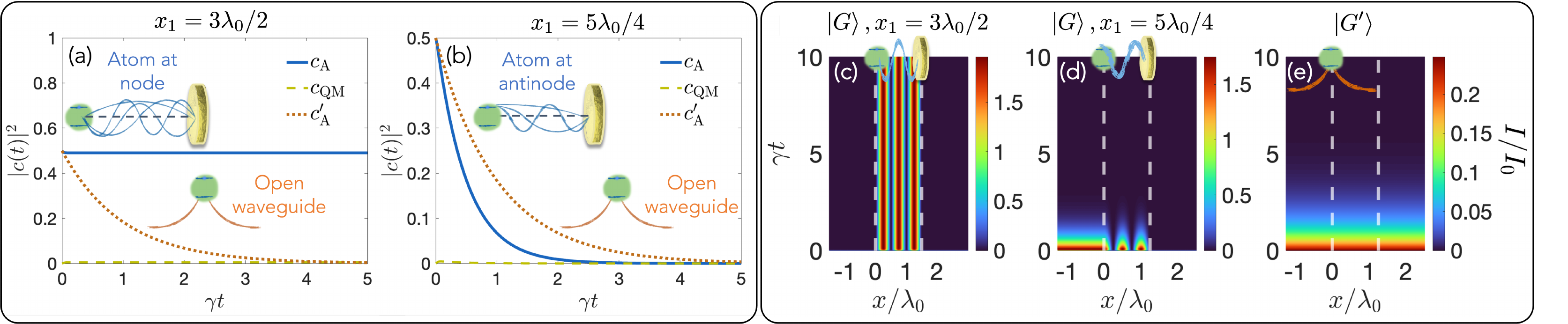}
    \caption{Preparing an atomic array with $ N = 100$ atoms initially in an equal superposition of the two ground states $ \frac{1}{\sqrt{2}} \bkt{\ket{G} + \ket{G'}}$,  the decay dynamics of the excited two-level atom is plotted for  (a) $ x_1 = 3\lambda_0 /2 $ and  (b) $ x_1 = 5\lambda_0 /4 $. The emitted intensity of the EM field, for a QM with $N = 100$,  (c) for the two-level atom-QM separation $ x_1 = 3\lambda_0 /2$ and state of QM $\ket{G}$ (field node at $x_A$) and  (d) $ x_1 = 5\lambda_0 /4$ ,  state of QM $\ket{G}$  (field antinode at $x_A$). (e) Intensity for QM in $\ket{G}'$ (independent of the Atom-QM separation).}
    \label{Fig:SingleQM2}
\end{figure*}

\textit{Three-level atomic array as a quantum mirror.}--- Let us consider an array of $N$ three-level atoms coupled to a waveguide, as shown in Fig.~\ref{Fig:SingleQM1}(a). The ground level $ \ket{g_n}$ (for the $n^\mr{th}$ atom in the array) is coupled to the excited level $\ket{e_n}$ with a resonant transition frequency $\omega_0$, while the level $\ket{g_n'}$ is decoupled from other levels. An initially excited two-level atom, also with a resonant frequency $ \omega_0$, is coupled to the waveguide at the position $x_A =0$. The interaction Hamiltonian that describes the interaction between the atoms and the guided field is given in the interaction picture as~\cite{SM}:
\eqn{\label{eq:Hint}&
\tilde H_\mr{int} =\non\\
&\sum_ k\sum_{n=  1}^N  \hbar g_k \hat a_k  \sbkt{\hat \sigma_n ^+ e ^{i k x_n } + \hat \sigma ^+ e^{i k x_A} }   e^{-i \bkt{\omega - \omega _0 }t } + \mr{H. c.},
}
where $ \hat \sigma_n ^+  \equiv \ket{e_n}\bra{g_n}$ is the raising operator for the $n^\mr{th}$ atom in the array and $ \hat \sigma^+ \equiv \ket{e}\bra{g}$ that for the two-level atom. The bosonic operators $ \hat a_k$ correspond to the EM field modes in the waveguide.  We consider the initial state of the system to be 
$\ket{\Psi(0) } = \bkt{c_A (0) \ket{G} +c_A' (0)\ket{G'}}\otimes\ket{e}_A\otimes \ket{\cbkt{0}}$, where the states $  \ket{G}\equiv \ket{g_1,g_2\dots g_n\dots g_N}$ and $\ket{G'}\equiv \ket{g'_1,g'_2\dots g'_n\dots g'_N}$ correspond to all the array atoms being in one ground level or another.

Considering that the interaction Hamiltonian preserves the total number of excitations in the atoms+field system, the state of the system at a later time $t$ in the single-excitation subspace is given by:
 \eqn{\label{Eq:psitQM}
\ket{\Psi (t) }=&\sbkt{\cbkt{c_\mr{QM} \hat \Sigma ^
+ + c_\mr{A}  \hat \sigma^+ + \sum_k c_{k} \hat a _k^\dagger}\ket{G} \right.\non\\
&\left.+ \cbkt{c_\mr{A}'  \hat \sigma^+ + \sum_k c'_{k} \hat a _k^\dagger  }\ket{G' }}\otimes\ket{g}\otimes\ket{\cbkt{0}},
}
where $c_A $ ($c_k $ ) and $ c_A'$ $(c_k') $  represent the excitation amplitudes for the two-level atom (field mode $k$), for the QM in state $\ket{G}$ and $ \ket{G'}$, respectively; $ c_\mr{QM}$ denotes the probability amplitude for there being an excitation in the QM. We have suppressed the explicit time dependence of the coefficients for brevity. $ \hat \Sigma ^+  = \frac{1}{\sqrt{N}}\sum_ {n = 1}^N (-1)^{ n+1 } \hat \sigma_n ^+ $ is the collective raising operator for  the array considering that the atoms are placed at half-wavelength ($\lambda_0/2 = \pi v/ \omega_0 $) apart from each other, $ v $ being the speed of light in the waveguide.  Such a Bragg mirror configuration, $|x_n - x_{n - 1}| = \lambda_0/2$, is imperative for the array  to behave cooperatively so that it can be described effectively as a single collective spin.  Eliminating the field modes yields the equations of motion for the atomic excitation amplitudes~\cite{SM}:
\eqn{
\der{c_\mr{QM}}{t}=&  - \frac{\gamma}{2} \bkt{N c_\mr{QM} +\sqrt{N}  e^{i k_0 x_1} c_\mr{A}}\\
\der{c_\mr{A}}{t} = & - \frac{\gamma}{2} \bkt{c_\mr{A}   + \sqrt{N}  e^{i k_0 x_1}c_\mr{QM}  };
\der{c'_\mr{A}}{t} =  - \frac{\gamma}{2}c'_\mr{A} , 
}
where we have  made the Born-Markov approximation, assuming a flat spectral density around the atomic resonance $(g_k \approx g_0 )$, and  defining $\gamma \equiv  2 \pi g_0 ^ 2$ as the spontaneous emission rate of individual atoms.  We observe that the QM couples to the field with a  cooperatively enhanced coupling $(\sim \sqrt{N} g_0)$.  Solving the above equations of motion, for  the QM in $ \ket{G'}$, the decay dynamics of the two-level atom is given by $c'_A (t) = c_A'(0) e^{- \frac{\gamma t}{2}} $, which corresponds to that in an open waveguide. For the QM in $ \ket{G}$, in the  limit of  $N\gg1$, the dynamics of the two-level atom simplifies to:
\eqn{c_{\mr{A}, \infty} (t) \approx c_\mr{A} (0)e^{i   \frac{\gamma t}{2} \sin\bkt{2 k_0 x_1} } e^{-  \frac{\gamma t}{2}\bkt{1 - \cos\bkt{2 k_0 x_1}} },} which corresponds to the decay of an excited two-level atom in front of a mirror~\cite{CookMilonni1987, MilonniKnight73}.   The accompanying coefficient for the  QM atoms to be excited $ (c_\mr{QM})$ scales as $\sim 1/\sqrt{N}$, vanishing in the large $N$ limit. We note that  the appropriate mirror-like boundary conditions emerge for large $N$, even when the QM atoms are negligibly excited. 

The spontaneous emission  dynamics for the two-level atom is illustrated in  Fig.\ref{Fig:SingleQM2}(a) and (b) as a function of the two-level atom-QM separation, assuming the initial state of the QM to be $ \frac{1}{\sqrt{2}}\bkt{\ket{G} + \ket{G'} }$. For the part of the total state in $ \ket{G'}$, the atom decays exponentially with a  rate $ \gamma$, as seen from the red dotted curves in Fig.\ref{Fig:SingleQM2}(a) and (b). The corresponding emitted intensity is plotted in Fig.\ref{Fig:SingleQM2}(e)~\cite{SM}.   For the part of the total state in $ \ket{G}$, the QM provides a mirror-like boundary condition, e.g., for a separation  $x_1 
 = p \lambda_0 /2$ ($ p \in \mathbb{N}$) between the two-level atom and the QM,  as the mirror atoms are excited thereby establishing a node of the field at $x_A=0$ (Fig.\ref{Fig:SingleQM2}(c)), inhibiting the subsequent decay of the two-level atom (blue solid curves in Fig.\ref{Fig:SingleQM2}(a)). Likewise, for $ x_1 = (2 p + 1 )\lambda_0 /4$ there is an antinode of the field at the position of the two-level atom (Fig.\ref{Fig:SingleQM2}(d)),  leading to a superradiant decay with $ 2\gamma$ (blue solid curves in Fig.\ref{Fig:SingleQM2}(b))~\cite{MilonniKnight73}.

\textit{Quantum erasure.}---From the time-evolved state in Eq.\eqref{Eq:psitQM}, as well as Fig.\ref{Fig:SingleQM2}(a) and (b), it is evident that the decay dynamics of the two-level atom is entangled with the the ground states of the QM. We now consider the possibility of `erasing' the state information of the QM by making a projective measurement on the QM atoms in a basis $\frac{1}{\sqrt{2}} \bkt{\ket{G}+e^{i \phi_M}\ket{G'}}$\footnote{While a multi-atom projective measurement can be challenging to implement in practice, we consider such a measurement as an idealized theoretical example. For example, one could apply a $\pi/2$-pulse to mix $\ket{G}$ and $\ket{G'}$,  and then post-select on $\ket{G}$. Alternatively, in experiments such as Ref. \cite{Srakaew2023}, a single qubit controls the whole state of the quantum mirror, making such projective measurements more accessible.
}~\cite{Scully1991}. Such a measurement gives the post-selected state at the measurement time $ t = t_M $:
 \vspace{-0.2 cm}
 \eqn{\label{Eq:psitQM2}
&\ket{\Psi (t_M) }=\non\\
&\frac{1}{2}\sbkt{ \cbkt{c_\mr{A}(t_M)  + e^{-i \phi_M}c'_\mr{A}(t_M)  }\hat \sigma^++ \sum_k \cbkt{\bkt{c_{k}(t_M) \right. \right. \right. \non\\
&\left.\left.\left.+e^{-i \phi_M} c'_{k}(t_M)} \hat a _k^\dagger }} \bkt{\ket{G} + e^{i \phi_M }\ket{G'}}\ket{g}\ket{\cbkt{0}}.
}
Having erased the which-state information for the QM, we see that the excitation amplitude of the two-level atom is now determined by a coherent superposition of $c_A(t_M)$ and $c'_A(t_M)$ corresponding to the two different boundary conditions. Starting with an initial state $\ket{\Psi(0) } = \frac{1}{\sqrt{2}}\bkt{ \ket{G} +e^{i \phi_S}\ket{G'}}\ket{e}\ket{\cbkt{0}}$, in the large $N$ limit, the excitation probability of the two-level atom right after the erasure becomes $ P_e(\phi_M) \equiv \frac{1}{2}\abs{c_A \bkt{t_M} + e^{- i \phi_M}c_A'\bkt{t_M}}^2$ such that:
\eqn{
&P_e(\phi_M)\approx \frac{e^{- \gamma t_M} }{4}\sbkt{ 1 + e^{\gamma t_M \cos (2 k_0 x_1)} \right.\non\\
&\left.+ 2 e^{\gamma t_M/2 \cos(2 k_0 x_1 ) } \cos\bkt{ \Delta \phi + \frac{\gamma t_M}{2} \sin \bkt{2 k_0 x_1}}},
}
where $ \Delta \phi \equiv \phi_M - \phi_S$. The first two terms in the above expression denote the independent excitation probabilities corresponding to the cases when the QM is in $ \ket{G'}$ and $ \ket{G}$ respectively; the third term arises from the interference between $ \ket{G'}$ and $ \ket{G}$. If $\ket{G}$ and $\ket{G'}$ have different energies, their respective free Hamiltonians will imprint a dynamical phase $\phi_S(t)=\delta t/\hbar $, where $\delta = \hbar (\omega_0 - \omega_0 ')$ represents the energy difference between the ground levels. Consequently, $\Delta \phi(t)$ becomes a function of time and $P_e$ will oscillate with $t_{M}$ as in a Ramsey interferometer of the two metastable states of the quantum mirror. Such  measurements can be experimentally  used to distinguish a coherent quantum superposition of two boundary conditions from a decohered statistical mixture.

 \begin{figure*}[t]
    \centering
\includegraphics[width = \textwidth]{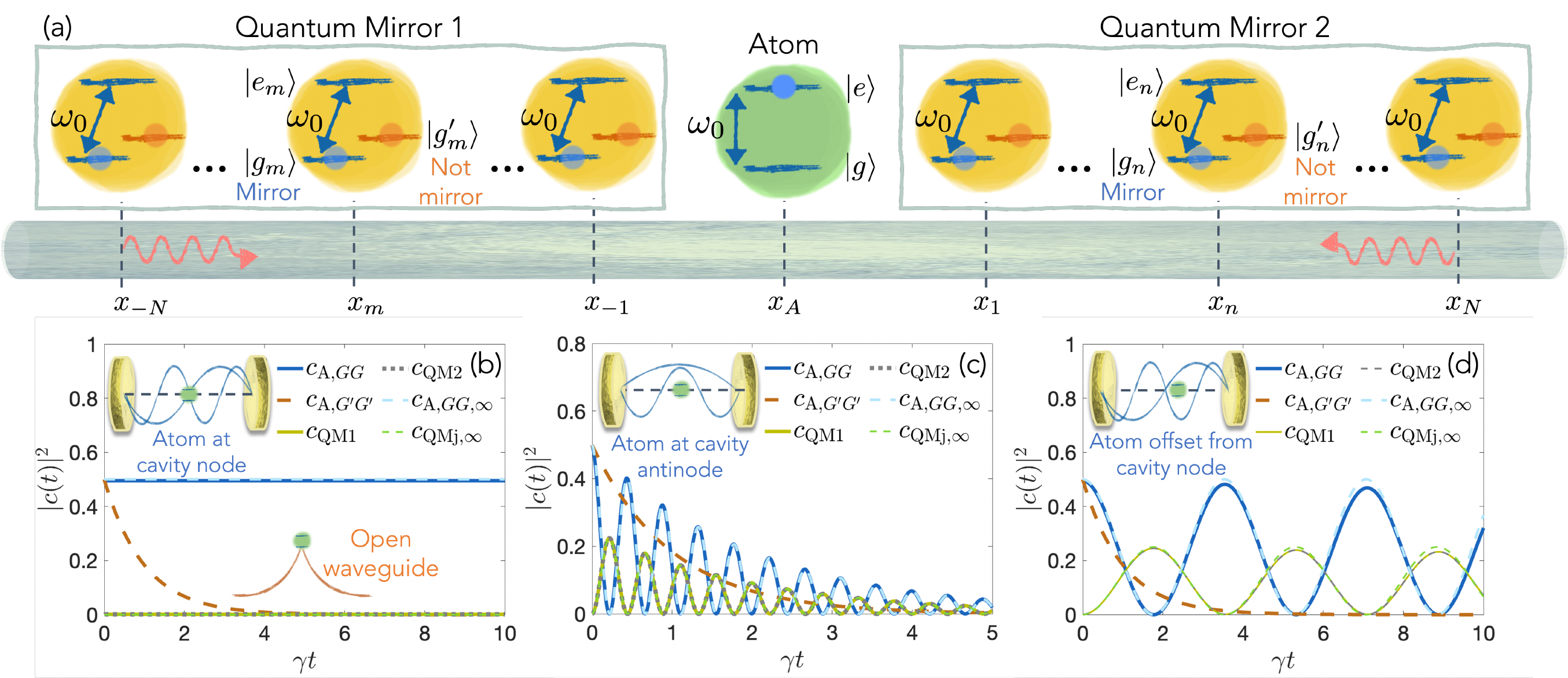}
    \caption{(a) Schematic representation of a two level atom in a quantum cavity. Spontaneous emission dynamics of the two-level atom in the quantum cavity for (b)$x_A = 0$, $ x_1 = 3 \lambda_0/2$, (c)$x_A = 0$, $ x_1 = 5 \lambda_0/4$ and (d) $x_A \approx 0.01 \lambda_0$, $ x_1 = 3 \lambda_0/2$. Each QM is composed of $N = 100 $ atoms. }
    \label{Fig:QCavity}
\end{figure*}

\textit{Spontaneous Emission in a Quantum Cavity.}---We now consider the spontaneous emission from a two-level atom inside a `quantum cavity' formed by two quantum mirrors, QM1 and QM2, as shown in Fig.~\ref{Fig:QCavity}(a). Each QM is composed of $N$ three-level atoms positioned at  $ x_j$, with $ j =\cbkt{1\dots N}$ $(j = \cbkt{- 1\dots - N})$ corresponding to QM1 (QM2). As in the case of a single QM, the spacing between the adjacent atoms in the array is $\lambda_0/2 $, such that each QM can be described in terms of a single collective spin.  Considering that the central two-level atom is initially excited and the atoms composing the QMs are all initially prepared either in  the state $ \ket{g_j}$ or $ \ket{g'_j}$, the state of the system at a later time $ t$ is given by~\cite{SM}:
\begin{widetext}
    \eqn{
&\ket{\Psi (t) } =\sbkt{\cbkt{c_{\mr{A} ,GG} \hat \sigma^+ + c_\mr{QM1} \hat \Sigma_{1}^+  + c_\mr{QM2}  \hat \Sigma_{2}^+  + \sum_k  c_{k,GG} \hat a _k ^\dagger }\ket{G_1G_2}+ \cbkt{c_{A,G'G'} \hat \sigma^+  + \sum_k  c_{k,G'G'}\hat a _k ^\dagger  }\ket{G'_1G'_2}\right.\non\\
&\left.+\cbkt{c_{\mr{A} ,GG'} \hat \sigma^+ + c'_\mr{QM1} \hat \Sigma_{1}^+ + \sum_k  c_{k,GG'} \hat a _k ^\dagger   }\ket{G_1G'_2}+\cbkt{c_{\mr{A} ,G'G} \hat \sigma^+ + c'_\mr{QM2} \hat \Sigma_{2}^+  + \sum_k  c_{k,G'G} \hat a _k ^\dagger }\ket{G'_1G_2}}\ket{g}\ket{\cbkt{0}},
}
\end{widetext}
where $ \ket{G_j} = \ket{g_1,g_2\dots  g_N}$ and $ \ket{G'_j} = \ket{g'_1,g'_2\dots  g'_N}$ indicates the collective state of QM$j$. The coefficients $ c_{\mr{A},G^{\cbkt{\prime}}G^{\cbkt{\prime}}}$ ($ c_{k,G^{\cbkt{\prime}}G^{\cbkt{\prime}}}$) denote the excitation amplitude of the two-level atom (field mode $k$), for the state of the QMs in $\ket{G_1^{\cbkt{\prime}}G_2^{\cbkt{\prime}}} $. Similarly the coefficients $ c^{\cbkt{\prime}}_\mr{QM1(2)}$ represent the probability amplitude for the QM1(2) to contain one collective excitation with the other QM in the state $ \ket{G_{2(1)}^{\cbkt{\prime}}}$.

While a detailed solution of the system dynamics  is presented in the SM~\cite{SM}, here we consider three example configurations as illustrated in Fig.~\ref{Fig:QCavity}. In each case, the system is prepared in an initial state $ \ket{\Psi (0) } = \frac{1}{\sqrt{2}}\ket{e}\otimes\bkt{ \ket{G_1G_2} + \ket{G_1'G_2'}}\otimes \ket{\cbkt{0}}$. For the part of the total state when the QMs are in $ \ket{G_1'G_2'} $, the two-level atom `sees' an open waveguide and decays exponentially (red dashed curves in Fig.\,\ref{Fig:QCavity}(b)--(d)). For the part of the state wherein the QMs are in the $\ket{G_1G_2}$ subspace,  the total system behaves effectively as  a two-level atom coupled to two collective spins corresponding to QM1 and QM2, each of which exhibits a cooperative coupling enhanced by a factor of $ \sqrt{N}$. The  dynamics strongly depends on the length of the cavity formed by the QMs and the position of the two-level atom therein, as elaborated below:
\begin{enumerate}
    \item {Atom at cavity node ($ x_A = 0 $, $x_1 = p \lambda_0 /2$, $ p \in \mathbb{N}$):   In this case the two-level atom is located at the node of the standing wave modes arising from the cavity-like boundary conditions, and thus stops decaying once the boundary conditions are established (solid blue curve in Fig.~\ref{Fig:QCavity}(b)). In the large $N$ limit,  it can be seen that $ c_{\mr{A},GG,\infty}(t)\approx c_{\mr{A},GG} (0)$ and  $ c_{\mr{QMj},\infty}(t)\propto 1/\sqrt{N} \rightarrow0$, as indicated by the blue-dashed and green-dashed curves in Fig.~\ref{Fig:QCavity}(b)~\cite{SM}. A similar condition arises when  $ 2x_1 < \lambda_0 /2$, such that there are no standing wave modes allowed within the cavity region~\cite{MilonniKnight73, Hulet1985}. }
    \item{Atom at cavity antinode ($ x_A = 0 $, $x_1 = (2p + 1) \lambda_0 /4$, $ p \in \mathbb{N}$): When the two-level  atom is located at the antinode of the standing wave mode in a resonant quantum cavity, it coherently exchanges excitations with the  atoms composing the QMs, as can be seen from Fig.~\ref{Fig:QCavity}(c).  This scenario is akin to the experimentally observed dynamics of superconducting qubits in  cavity QED system~\cite{Mirhosseini2019}.  In the large $N$ limit, the  dynamics of the two-level atom and the QMs is well-approximated by $c_{\mr{A }, GG,\infty} (t)  \approx c_{\mr{A},GG} (0) e^{- \gamma t/4}\cos \bkt{\sqrt{\frac{N}{2}}\gamma t}$ and $ c_\mr{QMj,\infty}\approx \sqrt{2}i c_{\mr{A},GG}(0) e^{- \gamma t /4} \sin \bkt{\sqrt{\frac{N}{2}}\gamma t}$, respectively~\cite{SM}, as seen from the blue and green dashed curves in Fig.~\ref{Fig:QCavity}(c). Remarkably, in this case the QMs do not admit an effective passive boundary condition description, with their excitation dynamics playing an active role in determining the system behavior even as $N\rightarrow\infty$. However, upon introducing a detuning between the atomic and cavity resonances, or displacing the two-level atom from the antinode, such a resonant excitation of the QMs disappears.  }
    
    \item{Atom near cavity node ($ 0<k_0 x_A \ll 1 $, $x_1 = p \lambda_0 /2$, $ p \in \mathbb{N}$): Being located close to the node  of the cavity, while the small offset from the node allows the two-level atom to still interact with the cavity modes, the coupling between the two-level atom and the open waveguide modes is minimized. We observe sustained Rabi oscillations-like dynamics between the two-level atom and the QM excitations, as depicted by the blue curve in Fig.~\ref{Fig:QCavity}(d). The dynamics of the two-level atom and those constituting  the QMs is approximated in the large $N$ limit by $ c_{\mr{A},GG,\infty }(t) \approx c_{\mr{A},GG }(0) \cos \bkt{k_0 x_A\sqrt{\frac{N}{2}}\gamma t}$ and $ c_\mr{QMj,\infty}(t)\approx-  \frac{i }{\sqrt{2}} \sin  \bkt{k_0 x_A\sqrt{\frac{N}{2}}\gamma t}$~\cite{SM} (blue and green dashed curves in Fig.~\ref{Fig:QCavity}(d)). As in case 2, the resonant excitation of the QM atoms vanishes in the presence of a detuning between the resonances of the two-level atom and  those forming the QMs, or that of the cavity.}
    
\end{enumerate}

\textit{Discussion.}---
We have demonstrated that an ordered array of waveguide-coupled $ \Lambda$-type three-level atoms,  prepared in a superposition of two ground states, can engender a superposition of two distinct boundary conditions, e.g., mirror-like and transparent, for the guided EM field modes. We examine the effect of such a quantum superposition of boundary conditions on the spontaneous emission of a two-level atom near such a \textit{quantum mirror} or QM, demonstrating that the corresponding atomic decay dynamics is entangled with the state of the QM. We show that making a measurement on the state of the atoms composing the QM can erase the `which-boundary-condition' information, resulting in an interference between the system dynamics  from the distinct boundary conditions.  We further analyze the spontaneous emission dynamics of a two-level atom inside a cavity formed by two QMs. Preparing the QMs in a  superposition of the two ground states, the atomic dynamics occurs in a superposition of having an open waveguide and the atom decaying inside a cavity. For specific arrangements  the atoms constituting the QMs can be resonantly excited, thereby precluding an effective static boundary condition description of the system.

Our results are pertinent to ongoing state-of-the-art experiments in waveguide QED~\cite{Sheremet23}, as well as those with free-space atomic arrays~\cite{Srakaew2023}. As was recently demonstrated in the experiment by Srakaew \textit{et al}, the optical response of a subwavelength 2D array of Rydberg atoms can be switched from mirror-like to transparent via an ancilla atom. Preparing such a system in a superposition can potentially result in a quantum mirror-like boundary condition, which can be tested by measuring the system in a basis that erases the `which-state' information on the ancilla, thereby leading to an interference between the two boundary conditions.

From the perspective of applications in quantum information processing, similar subwavelength 2D arrays of atoms  have also been proposed for generation of photonic Greenberger-Horne-Zeilinger (GHZ) states~\cite{Bekenstein20}. Here we start with a GHZ state of atomic arrays coupled to a waveguide and obtain the subsequent dynamics of the system. Future works will explore  time-reversed renditions of the present dynamics wherein one can potentially herald an entangled state by starting with an initially unentangled state and making appropriate measurements on the system.

Finally, quantization of EM field in the presence of  boundary conditions that can be in a quantum superposition or quantum correlated states, is as yet an open problem. Such a quantization scheme can be studied as an extension of the previous works on microscopic models of boundary conditions~\cite{CookMilonni1987}, assuming the induced polarization in a medium to be in a macroscopic quantum superposition. This will  open a new direction of exploring QED phenomena in the presence of `quantum boundary conditions'.

\textit{Acknowledgments.}--- 
We warmly acknowledge insightful discussions with Elizabeth A. Goldschmidt, Hakan T\"ureci, Peter W. Milonni and Pierre Meystre.
K.S. acknowledges support from the National Science Foundation under Grant No. PHY-2309341, and by the John Templeton Foundation under Award No. 62422.  P.S. is a CIFAR Azrieli Global Scholar in the Quantum Information Science Program. This research was supported in part by grant NSF PHY-1748958 to the Kavli Institute for Theoretical Physics (KITP).
\bibliography{AtomicMirror.bib}

\clearpage
\onecolumngrid
\begin{center}
	
	\newcommand{\beginsupplement}{
		\setcounter{table}{0}
		\renewcommand{\thetable}{S\arabic{table}}
		\setcounter{figure}{0}
		\renewcommand{\thefigure}{S\arabic{figure}}
		\setcounter{equation}{0}
		\renewcommand{\theequation}{S\arabic{equation}}
	}
	\beginsupplement
	
	\textbf{\large Supplemental Material for ``Spontaneous emission in the presence of Quantum Mirrors''}
\end{center}

\maketitle
\makeatletter
\renewcommand{\theequation}{S\arabic{equation}}
\renewcommand{\thefigure}{S\arabic{figure}}
\renewcommand{\bibnumfmt}[1]{[S#1]}
\renewcommand{\citenumfont}[1]{S#1}

\vspace{-0.5 cm}

\section{Single quantum mirror}

The total Hamiltonian for a two-level atom near a three-level  atom array (QM) is given by:
\eqn{
H = H_A + H_F + H_\mr{int},
}
where
\eqn{H_A = &\hbar \omega_0 \ket{e} \bra{e} + \sum_{n  = 1}^N \hbar \omega_0 \ket{e_n} \bra{e_n}    ,
}
is the atomic Hamiltonian, $ H_F$ is the field Hamiltonian and the interaction Hamiltonian in the interaction picture is as given in the main text.

Let us consider the state at time $ t$ as:

\eqn{\ket{\Psi (t) }=& \sbkt{\sum _{n 
 = 1}^N\cbkt{c_{\mr{QM}, n}(t) \hat \sigma_n ^
+ + c_\mr{A} (t) \hat \sigma^+ + \sum_k c_{k}(t) \hat a _k^\dagger  }\ket{G} + \cbkt{c_\mr{A}' (t) \hat \sigma^+ + \sum_k c'_{k}(t) \hat a _k^\dagger  }\ket{G' }}\ket{g}\ket{\cbkt{0}}.
}
We obtain the evolution of the atomic and field excitation amplitudes as follows:

\eqn{
\der{c_{\mr{QM}, n}}{t} =& -i \sum_k g_kc_{k}(t) e^{i k x_n}  e^{-i \bkt{\omega - \omega_0 }t }\\
\der{c_\mr{A}}{t} =&  -i \sum _k g_kc_{k} (t) e^{-i (\omega - \omega_0 )t};
\quad\der{c'_\mr{A}}{t} = -i \sum _k g_kc'_{k} (t) e^{-i (\omega - \omega_0)t}\\
\label{Eq:cak}
\der{c_{k}}{t} = & -i g_k \sbkt{c_\mr{A}(t)e^{i (\omega - \omega_0 )t} + \sum_{n 
 = 1}^Nc_{\mr{QM},n}(t)e^{-i k x_n}e^{i (\omega - \omega_0 )t}};\quad \der{c'_{k}}{t} =  -i g_k c'_\mr{A}(t)e^{i (\omega - \omega_0 )t} 
}
Substituting the field excitation amplitudes in the atomic equations of motion, yields:

\eqn{
\der{c_{\mr{QM},n}}{t}  \approx & - \frac{\gamma}{2}\sbkt{c_\mr{A}\bkt{t  }e ^{i \omega_0 x_n/v}+\sum_{\nu = 1}^N c_{\mr{QM},\nu}\bkt{t } e ^{i \omega_0 \abs{x_n - x_\nu}}}\\
\der{c_\mr{A}}{t} \approx& - \frac{\gamma}{2} \sbkt{c_\mr{A}\bkt{t }+\sum_{n = 1}^N c_{\mr{QM},n }\bkt{t  }e^{i \omega_0 x_n /v} }\\
\der{c'_\mr{A}}{t} \approx&  - \frac{\gamma }{2} c_{G'}^{e,0}(t),
}
where we have assumed  that $ g_k \approx g_0 $ for the field modes around the atomic resonance, and the spontaneous emission rate of an atom into a waveguide is defined as $ \gamma \approx 2 \pi \abs{g_0} ^2$.

Taking Laplace transform of the above equations of motion as $ \tilde c_j  (s) \equiv \int_0 ^\infty \dd t c_j (t) e^{- s t} $, we get:

\eqn{
\bkt{s + \frac{\gamma}{2}}  c_{\mr{QM},n} ( s) + \frac{\gamma}{2}\tilde c_\mr{A}\bkt{s }e ^{i \omega_0 x_n/v}+ \frac{\gamma}{2}\sum_{\nu\neq n } \tilde c_{\mr{QM},\nu }\bkt{s}e ^{i \omega_0 \abs{x_n - x_\nu}/v}= &c_{\mr{QM},n}(0) \\
\bkt{ s + \frac{\gamma}{2}} \tilde c_\mr{A} ( s) + \frac{\gamma}{2}\sum_{n = 1}^N \tilde c_{\mr{QM},n }\bkt{ s }e ^{i \omega_0 x_n/v}= &c_\mr{A} (0) \\
\bkt{ s + \frac{\gamma}{2}} \tilde c'_\mr{A} ( s)  =  & c'_{A}(0).
}

One can define the collective excitation amplitude for the mirror atoms as:

\eqn{c_\mr{QM}(t)  \equiv & \frac{1}{\sqrt{N}}\sum_{n  = 1}^N c_{\mr{QM},n } (t)(-1)^{n- 1},}
where  we have assumed that

\eqn{
x_n = x_1 + (n - 1) \lambda_0 /2.
}

The Laplace equations of motion thus simplify to:

\eqn{\label{Eq:tildecqm}
\bkt{ s + \frac{N\gamma}{2}} \tilde c_\mr{QM} (s) + \frac{\sqrt{N}\gamma}{2}\tilde c_\mr{A}\bkt{ s}e ^{i k_0 x_1}= &c_{\mr{QM}}(0) \\
\label{Eq:tildeca}
\bkt{s + \frac{\gamma}{2}} \tilde c_\mr{A} ( s) + \frac{\sqrt{N}\gamma}{2}\tilde c_\mr{QM }\bkt{ s}e ^{i k_0 x_1}= &c_\mr{A} (0) \\
\bkt{s + \frac{\gamma}{2}} \tilde c'_{A} ( s)  =  & c'_{A}(0).
}

The equations of motion Eq.(3--4) in the main text correspond to the inverse Laplace transform of the above.

Assuming that the atoms in the QM are initially unexcited ($ c_\mr{QM}(0) = 0 $), we obtain:

\eqn{\label{eq:cas}
\tilde c_\mr{A}(s) = &\frac{s + \frac{N\gamma}{2}}{\bkt{s + \frac{\gamma}{2}}\bkt{s + \frac{N\gamma}{2}} - \frac{N\gamma^2}{4} e^{2 i k_0 x_1}}c_\mr{A} (0)\\
\label{eq:cqms}
\tilde c_\mr{QM}(s) = &- \frac{\sqrt{N}\gamma}{2}e^{i k_0 x_1}\frac{1}{\bkt{s + \frac{\gamma}{2}}\bkt{s + \frac{N\gamma}{2}} - \frac{N\gamma^2}{4} e^{2 i k_0 x_1}}c_\mr{A} (0)
}

The poles of the above equation are given by $ \gamma_\pm  \equiv - \frac{\gamma}{2}\sbkt{ \frac{N + 1} {2}\pm \sqrt{\frac{ (N + 1)^2}{4} + N e^{2 i k_0 x_1} } }$. This yields the time dynamics of the two-level atom as:
\eqn{
c_\mr{A} (t) = \frac{\gamma_+ + \frac{N \gamma}{2}}{\gamma_+ - \gamma_-}e^{\gamma_+ t} -\frac{\gamma_- + \frac{N \gamma}{2}}{\gamma_+ - \gamma_-}e^{\gamma_- t} \\
c_\mr{QM} (t) = - \frac{\sqrt{N}\gamma}{2\bkt{\gamma_+ - \gamma_-}}e^{i k_0 x_1}\bkt{e^{\gamma_+ t}- e^{\gamma_- t}}. 
}

\subsection{Large $N $ limit}

For $ N\gg 1$, one obtains from  Eq.\,\eqref{eq:cas}:

\eqn{\tilde c_\mr{A} ( s ) \approx  \frac{c_\mr{A}(0) }{ s + \frac{\gamma}{2} \bkt{1 - e^{2 i k_0 x_1}}} \implies 
c_\mr{A} (t) \approx e^{- \frac{\gamma t}{2} \bkt{1 + e^{2 i k_0 x_1}}},
}
which corresponds to the decay of an excited atom in front of a mirror~\cite{CookMilonni1987}. From Eq.\,\eqref{eq:cqms} it can be seen that the collective excitation amplitude for the QM atoms scales as $c_\mr{QM} \sim 1/\sqrt{N}$, which vanishes in the large $N $ limit.

\subsection{Field dynamics}

The intensity of the field emitted by the atoms as a function of position and time can be evaluated as $I\bkt{x,t} = \frac{\epsilon_0 c} {2}\avg{\Psi\bkt{t}\vert\hat{E}^\dagger\bkt{x,t} \hat E\bkt{x,t}\vert\Psi\bkt{t}} $, where $\hat{E}\bkt{x,t} = \sum_k\,\mc{E}_k \hat{a}_k e^{i (k x- \omega t)}$ is the electric field operator at position $x$ and time $t$. More explicitly, we obtain

\eqn{\label{eq:int}
I (x,t)/I_0 =& \bra{\Psi (t) }\sbkt{\sum_{k_1}\hat a_{k_1}^\dagger e^{-i (k_1 x - \omega_1 t)}}\sbkt{\sum_{k_2} \hat a_{k_2} e^{i (k_2 x - \omega_2 t)} }\ket{\Psi (t) }\\
= & \sum_{k_1} \sum_{k_2} \sbkt{ \bra{G}\bra{g}c_{k_1}^\ast e^{-i k_1 x} +  \bra{G'}\bra{g}(c'_{k_1})^\ast e^{-i k_1 x} } \sbkt{ c_{k_2} e^{i k_2 x} \ket{G}\ket{g}+  c'_{k_2} e^{i k_2 x}   \ket{G'}\ket{g}} e^{i (\omega_1 - \omega_2) t}\\
 = & \abs{\sum_{k}c_{k}(t) e^{i (k x - \omega t)}}^2 + \abs{\sum_{k}c'_{k}(t) e^{i (k x- \omega t)}}^2 }

 We substitute the field excitation amplitudes in terms of the atomic excitation amplitudes using Eq.\eqref{Eq:cak} and  separating the positive and negative wavevectors $k$ explicitly.:
 \eqn{
I/I_0  =&  \abs{\int_0 ^t\dd \tau e^{-i \omega_0 \tau } \int _0 ^\infty 
\dd\omega e^{-i \omega (t- \tau)}\sbkt{\cbkt{c_\mr{A}\bkt{\tau } + \sum _n c_{\mr{QM},n}\bkt{\tau }e^{-i k x_n} } e^{ i k x}\right.\right.\non\\
&\left.\left.+\cbkt{  c_\mr{A}\bkt{\tau }+\sum_n c_{\mr{QM},n }\bkt{\tau }e^{i k x_n}}  e^{ -i k x} }}^2+ \abs{\int_0 ^t\dd \tau e^{-i \omega_0 \tau } \int_0 ^\infty \dd\omega  e^{-i \omega (t -\tau)}  \sbkt{c'_\mr{A}(\tau) e^{ i k x} + c'_\mr{A}(\tau) e^{ -i k x} }}^2 \\
  =&  \abs{\sbkt{e^{i \omega_0  x/v} c_\mr{A}\bkt{t - x/v}\sbkt{\Theta \bkt{t -  x/v} -\Theta \bkt{ -x/v}   }+ e^{-i \omega_0  x/v} c_\mr{A}\bkt{t + x/v}\sbkt{\Theta \bkt{t  +  x/v} - \Theta \bkt{x/v}}\right.\right.\non\\
 &\left.\left. +\sum_n  e^{i \omega_0   (x - x_n )/v }c_{\mr{QM},n }\bkt{t-  (x - x_n)/v }\sbkt{\Theta \bkt{t -  (x - x_n)/v} - \Theta \bkt{ -(x -x_n)/v}} \right.\right.\non\\
  &  \left.\left.+\sum_n e^{-i \omega_0  (x -x_n)/v )}c_{\mr{QM},n }\bkt{t+ (x - x_n )/v  }\sbkt{\Theta \bkt{t+   (x-x_n )/v} -\Theta \bkt{    (x -x_n )/v}  } }}^2\non\\
 &+ \abs{ e^{-i \omega_0  x/v} c'_\mr{A}(t + x/v) \sbkt{\Theta \bkt{t + x/v} - \Theta \bkt{ x/v}}  +  e^{i \omega_0   x/v } c'_{A}(t - x/v) \sbkt{\Theta \bkt{t - x/v} - \Theta \bkt{ -x/v}}  }^2 .
}
We note that the total intensity emitted by the atoms is an incoherent sum of that emitted from the QM atoms in  $ \ket{G}$ and $ \ket{G'}$. In the limit where the QM atoms have a separation that is small comparable to the coherence length of the emitted photon $ \gamma\abs{x_1 - x_{N}}/c\ll1 $, the time delay effects between the different $ c_{\mr{QM}, n}$ coefficients can be ignored such that $c_\mr{QM, n}(t \pm (x - x_n)/v)\approx c_\mr{QM, n}(t \pm (x - x_1)/v)$~\cite{Sinha20a, Sinha20b}. Thus in the absence of time-delay effects the above expression for the intensity can be simplified  to:

\eqn{
I/I_0 \approx &\abs{\sbkt{e^{i \omega_0  x/v} c_\mr{A}\bkt{t - x/v}\sbkt{\Theta \bkt{t -  x/v} -\Theta \bkt{ -x/v}   }+ e^{-i \omega_0  x/v} c_\mr{A}\bkt{t + x/v}\sbkt{\Theta \bkt{t  +  x/v} - \Theta \bkt{x/v}}\right.\right.\non\\
 &\left.\left. +e^{i \omega_0   (x - x_1)/v }c_{\mr{QM} }\bkt{t-  (x - x_1)/v }\sbkt{\Theta \bkt{t -  (x - x_1)/v} - \Theta \bkt{ -(x -x_1)/v}} \right.\right.\non\\
  &  \left.\left.+ e^{-i \omega_0  (x -x_1)/v }c_{\mr{QM}}\bkt{t+ (x - x_1)/v  }\sbkt{\Theta \bkt{t+   (x-x_1 )/v} -\Theta \bkt{    (x -x_1)/v}  } }}^2\non\\
 &+ \abs{ e^{-i \omega_0  x/v} c'_\mr{A}(t + x/v) \sbkt{\Theta \bkt{t + x/v} - \Theta \bkt{ x/v}}  +  e^{i \omega_0   x/v } c'_{A}(t - x/v) \sbkt{\Theta \bkt{t - x/v} - \Theta \bkt{ -x/v}}  }^2 .
}

\section{Quantum Cavity}

In this section we consider an atomic array as a mirror. For the sake of simplicity the initial state of the total system is assumed to be:

\eqn{
\ket{\Psi (0)} = \ket{e} \otimes \bkt{{c}_{\mr{A},GG}(0)\ket{G_1}\ket{G_2} + {c}_{\mr{A},GG'}(0)\ket{G_1 }\ket{G'_2 }+ {c}_{\mr{A},G'G}(0)\ket{G'_1}\ket{G_2} + {c}_{\mr{A},G'G'}(0)\ket{G'_1} \ket{G'_2} }\ket{\cbkt{0}}
}
where we consider  all of the atoms in QM1 and QM2 to be initially in states $ \ket{G_{1,2}} = \ket{gg...g}_{1,2}$ or $ \ket{G'_{1,2}} = \ket{g'g'...g'}_{1,2}$. 

The total Hamiltonian for this system is:
\eqn{
H = H_A + H_F + H_\mr{int}
}
where 
\eqn{H_A = \hbar \omega_0\ket{e}\bra{e} + \sum_{m = -1}^{-N} \hbar \omega_0 \ket{e_m}\bra{e_m} + \sum_{n  =1}^N \hbar \omega_0 \ket{e_n}\bra{e_n} .}

Here $m $ and $ n $ are the index for the mirror atoms in QM1 on the left, and QM2 on the right.

The interaction Hamiltonian in the interaction picture becomes
\eqn{
\tilde H_\mr{int} = \sum_k \hbar g_{k}\hat a_k \bkt{  \sum_{m = -1}^{-N} \hat \sigma_m^+   e^{i k x_m} + \sum_{n = 1}^N \hat \sigma_n^+    e^{i k x_n} + \sum_k \hat \sigma^+   e^{i k x_A}}e^{-i (\omega - \omega_0 )t } + \mr{H. c.}
}
With the above interaction, the atomic state at any time is given by:

   \eqn{
\ket{\Psi (t) } = &\sbkt{\cbkt{c_{\mr{A} ,GG}(t) \hat \sigma^+ +\sum_{m = -1}^{-N}  c_{e_mG}(t) \hat \sigma_{m}^+  +\sum_{n = 1}^Nc_{Ge_n} (t) \hat \sigma_{n}^+  + \sum_k  c_{k,GG}(t) \hat a _k ^\dagger   }\ket{G_1}\ket{G_2} \right.\non\\
&\left.+\cbkt{c_{\mr{A} ,GG'}(t) \hat \sigma^+ + \sum_{m = -1}^{-N}c_{e_mG'}(t) \hat \sigma_{m}^+ + \sum_k  c_{k,GG'}(t) \hat a _k ^\dagger  }\ket{G_1}\ket{G'_2}\right.\non\\
&\left.+\cbkt{c_{\mr{A} ,G'G}(t) \hat \sigma^+ +\sum_{n = 1}^N   c_{G'e_n}(t) \hat \sigma_{n}^+  + \sum_k  c_{k,G'G}(t) \hat a _k ^\dagger   }\ket{G'_1}\ket{G_2}\right.\non\\
&\left.+ \cbkt{c_{\mr{A},G'G'}(t) \hat \sigma^+  + \sum_k  c_{k,G'G'}(t)\hat a _k ^\dagger  }\ket{G'_1}\ket{G'_2}}\ket{g}\ket{\cbkt{0}}.
}

The equations of motion for the atomic excitation amplitudes are:

\eqn{
\der{c_{\mr{A},GG} (t)}{t} = &-i  \sum_k g_k c_{k,GG} (t) e^{i k x_A} e^{-i (\omega - \omega_0 )t }; \quad
\der{c_{\mr{A},GG'} (t)}{t} = -i  \sum_k g_k c_{k,GG'} (t)  e^{i k x_A}e^{-i (\omega - \omega_0 )t }\\
\der{c_{\mr{A},G'G} (t)}{t} = &-i  \sum_k g_k c_{k,G'G} (t) e^{i k x_A} e^{-i (\omega - \omega_0 )t };\quad
\der{c_{\mr{A},G' G'}(t)}{t} = -i  \sum_k g_k c_{k,G' G'} (t)  e^{i k x_A}e^{-i (\omega - \omega_0 )t }
}
Similarly, we have  the equations of motion for the excitation amplitudes of the atoms composing the quantum mirrors:
\eqn{
\der{c_{e_mG} (t) }{t} = &-i  \sum_k g_{k} c_{k,G G}(t)e^{i k x_m }  e^{-i (\omega - \omega_0 )t }; \quad \der{c_{e_{m} G'} (t) }{t} = -i  \sum_k g_{k} c_{k, G G'}(t)e^{i k x_m }  e^{-i (\omega - \omega_0 )t }\\
\der{c_{ Ge_{n}} (t) }{t} = &-i  \sum_k g_{k} c_{ k, GG}(t)e^{i k x_n}  e^{-i (\omega - \omega_0)t };\quad
\der{c_{ G'e_{n}} (t) }{t} = -i  \sum_k g_{k} c_{k, G'G}(t)e^{i k x_n}  e^{-i (\omega - \omega_0 )t }
}
For the field excitation amplitudes:
\eqn{
\der{c_{k, GG}(t)}{t} = & -i\sbkt{ g_k c_{\mr{A},GG} (t)e^{-i k x_A} +\sum_{m = -1}^{-N} g_{k} c_{e_mG} (t) e^{-i k x_m } + \sum_{n = 1}^N g_{k} c_{Ge_n} (t) e^{-i k x_n }}e^{i (\omega - \omega_0 )t}\\
\der{c_{k,GG'}(t)}{t} = & -i \sbkt{g_k c_{\mr{A},GG'}(t) e^{-i k x_A} + \sum_{m = -1}^{-N} g_{k} c_{e_mG'} (t) e^{-i k x_m }}e^{i (\omega - \omega_0 )t}\\
\der{c_{k,G'G}(t)}{t} = & -i \sbkt{g_k c_{\mr{A},G'G} (t) e^{-i k x_A} + \sum_{n = 1}^Ng_{k} c_{G'e_n} (t) e^{-i k x_m } }e^{i (\omega - \omega_0 )t}\\
\der{c_{k,G'G'}(t)}{t} = & -i g_k c_{\mr{A},G'G'}(t)e^{-i k x_A}e^{i (\omega - \omega_0)t}
}

Substituting the field amplitudes in the atomic equations of motion, and  making the Born-Markov approximation one obtains:

\eqn{\label{eq:caggt}
\der{c_{\mr{A},GG}(t)}{t} \approx& -\frac{\gamma}{2} \sbkt{  c_{\mr{A},GG} (t) + \sum_{m = -1}^{-N} c_{e_mG} (t) e^{-i  \omega_1 (x_m - x_A)/v} + \sum_{n = 1}^N c_{Ge_n} (t)e^{i \omega_2 (x_n-x_A)/v} }\\
\label{eq:cagg't}
\der{c_{\mr{A},GG'}(t)}{t} 
\approx &-\frac{\gamma}{2} \sbkt{c_{\mr{A},GG'}(t)+ \sum_{m = -1}^{-N} c_{e_mG'} (t)e^{-i \omega_1 (x_m - x_A)/v} }\\
\label{eq:cag'gt}
\der{c_{\mr{A},G' G}(t)}{t} \approx& - \frac{\gamma}{2}\sbkt{ c_{\mr{A},G'G} (t) +\sum_{n = 1}^N c_{G' e_n} (t)e^{i \omega_2 (x_n - x_A) /v } }\\
\label{eq:cag'g't}
\der{c_{\mr{A},G' G'} (t)}{t} \approx &- \frac{\gamma}{2} c_{\mr{A },G'G'} (t)\\
\label{eq:cegt}
 \der{c_{e_{m} G} (t) }{t}  \approx & -\frac{\gamma}{2}\sbkt{   \sum_{\mu = -1}^{-N} c_{e_\mu G} (t) e^{i \omega_1 \abs{x_m- x_\mu} /v}+ c_{\mr{A},GG} (t)e^{-i \omega_A (x_m- x_A)/v}+ \sum_{n = 1}^Nc_{Ge_n} (t) e^{i \omega_2 (x_n - x_m)/v} }\\
 \label{eq:ceg't}
 \der{c_{e_{m}G'} (t) }{t} 
\approx & - \frac{\gamma}{2}\sbkt{  \sum_{\mu = -1}^{-N}  c_{e_\mu G'} (t) e^{i \omega_1  \abs{x_m - x_\mu}/v }+ c_{\mr{A},GG'} (t) e^{- i\omega_A  (x_m - x_A)/v } } \\
\label{eq:cget}
 \der{c_{ Ge_{n}} (t) }{t} \approx & -\frac{\gamma}{2} \sbkt{ c_{\mr{A},GG} (t) e^{i \omega_A  (x_n - x_A)/v}+ \sum_{m = -1}^{-N}  c_{e_mG} (t)  e^{i \omega_1 ( x_n-x_m)/v}+ \sum_{\nu = 1}^N c_{Ge_\nu} (t) e^{i \omega_2 \abs{ x_n-x_\nu}/v}}\\
 \label{eq:cg'et}
 \der{c_{ G'e_{n}} (t) }{t} \approx & -\frac{\gamma}{2}   \sbkt{ c_{\mr{A}, G'G} (t)e^{i \omega_A  (x_n - x_A)/v} + \sum_{\nu = 1}^N  c_{G'e_\nu} (t) e^{i \omega_2  \abs{x_n - x_\nu}/v} }.
}

We have assumed  the spontaneous emission rate to be $ \gamma \approx 2 \pi |g\bkt{\omega_{0}}|^2$  for a sufficiently flat spectral density of the field.  We now define the collective excitation amplitudes for the QMs:

\eqn{
c_\mr{QM1}(t)  \equiv & \frac{1}{\sqrt{N}}\sum_{m  = 1}^N c_{e_m G} (t)(-1)^{m- 1};\quad c'_\mr{QM1}(t)  \equiv  \frac{1}{\sqrt{N}}\sum_{m  = 1}^N c_{e_m  G'} (t) (-1)^{m - 1}\\
c_\mr{QM2}(t)  \equiv&  \frac{1}{\sqrt{N}}\sum_{n  = 1}^N c_{G e_n} (t) (-1)^{n - 1};\quad
c'_\mr{QM2}(t)  \equiv \frac{1}{\sqrt{N}}\sum_{n  = 1}^N c_{G'e_n} (t) (-1)^{n - 1}
}

Note that the position of the $ n ^\mr{th}$ atom in mirror 2 is 
\eqn{x_n =&  x_1 + (n - 1)\lambda_0 /2\\
\implies  e^{i k_0 x_n } = & e^{i k_0  x_1 } e^{i \pi  (n - 1)} = e^{i k_0  x_1 } (-1)^{ (n - 1)} 
} Similarly, the position of the $ m^\mr{th}$ atom in mirror 1 is: 
 \eqn{ x_m = &  x_{-1} - (m - 1)\lambda_0 /2\\
\implies  e^{- i k_0 x_m } = & e^{-i k_0 x_{-1} } e^{i \pi  (m - 1)} = e^{i k_0  x_{1} } (-1)^{ (m - 1)},
} 
where we have assumed, without loss of generality, that $ x_{-1} = - x_1$.

Taking the Laplace transform of  the atomic equations of motion (Eq.\eqref{eq:caggt}--\eqref{eq:cg'et}) in terms of the collective mirror excitation amplitudes, we obtain:
\eqn{\label{eq:cagg}
\bkt{s + \frac{\gamma}{2}}\tilde c_{\mr{A},GG} (s) + \frac{\sqrt{N}\gamma} {2} e^{i k_0( x_{1} + x_A)}\tilde c_\mr{QM1}(s)   +\frac{\sqrt{N}\gamma}{2}e^{i k_0 (x_1 - x_A)} \tilde  c_\mr{QM2}(s) = c_{\mr{A},GG} (0)\\
\label{eq:cagg'}
\bkt{s +  \frac{\gamma}{2}} \tilde c_{\mr{A},GG'} (s)  + \frac{\sqrt{N}\gamma}{2}e^{i k_0 (x_{1} + x_A)}\tilde c'_\mr{QM1} (s)= c_{\mr{A},GG'} (0) \\
\label{eq:cag'g}
\bkt{s + \frac{\gamma}{2}}\tilde c_{\mr{A},G'G} (s)  + \frac{\sqrt{N}\gamma}{2}e^{i k_0 (x_1 - x_A) }\tilde c'_\mr{QM2} (s) = c_{\mr{A},G'G} (0) \\
\label{eq:cag'g'}
\bkt{s + \frac{\gamma}{2}}\tilde c_{\mr{A},G' G'} (s) = c_{\mr{A},G' G'}(0) \\
\label{eq:ceg}
\bkt{s + \frac{N \gamma}{2}} \tilde c_\mr{QM1} (s) + \frac{\sqrt{N}\gamma}{2} e^{i k_0 (x_{1} + x_A)}\tilde c_{\mr{A},GG} (s)+ \frac{N \gamma}{2} e^{2i k_0 x_1 }\tilde c_\mr{QM2} (s ) 
=  c_\mr{QM1}(0)\\
\label{eq:ceg'}
\bkt{s + \frac{N\gamma}{2}} \tilde c'_\mr{QM1} (s) +\frac{\sqrt{N}\gamma}{2} e^{ ik_0 (x_{1} + x_A)}\tilde c_{\mr{A},GG'} (s)  = c'_\mr{QM1} (0)\\
\label{eq:cge}
\bkt{s + \frac{N\gamma}{2}}\tilde c_\mr{QM2} (s) +\frac{\sqrt{N}\gamma}{2} e^{i k_0 (x_1 - x_A)} \tilde  c_{\mr{A},GG} (s)   + \frac{N \gamma}{2}e^{2i k_0 x_1 } \tilde c_\mr{QM1} (s )   
=c_\mr{QM2}(0)  \\
\label{eq:cg'e}
\bkt{s + \frac{N\gamma}{2}}\tilde c'_\mr{QM2}(s)  +\frac{\sqrt{N}\gamma}{2}e^{i k_0 (x_1 - x_A)}   \tilde c_{\mr{A},G'G}(s)= c'_\mr{QM2}(0),
}
Each QM acts as an effective collective spin with a cooperative coupling of $ \sqrt{N} \gamma$ to the waveguide. Eq. \eqref{eq:cag'g'} can be immediately solved to obtain an exponential decay for $ c_{\mr{A},G'G'} (t) = c_{\mr{A},G'G'} (0)e^{-\gamma t/2}$.

The remaining equations can be expressed in matrix notation as:

\eqn{\dbar M (\tilde s) \underbrace{\bkt{\begin{array}{c}
      \tilde c_{\mr{A},GG}(\tilde s ) \\
      \tilde c_{\mr{A},GG'}(\tilde s)\\
      \tilde c_{\mr{A},G'G}(\tilde s)\\
      \tilde c_{\mr{QM1}}(\tilde s)\\
      \tilde c'_{\mr{QM1}}(\tilde s)\\
      \tilde c_{\mr{QM2}}(\tilde s)\\
      \tilde c'_{\mr{QM2}}(\tilde s)
\end{array}}}_{ \equiv \tilde{\mb{C}}(\tilde s) } = \frac{1}{\gamma}\underbrace{\bkt{\begin{array}{c}
       c_{\mr{A},GG} (0)\\
       c_{\mr{A},GG'}(0)\\
       c_{\mr{A},G'G} (0)\\
      c_{\mr{QM1}} (0)\\
       c'_{\mr{QM1}} (0)\\
       c_{\mr{QM2}} (0)\\
       c'_{\mr{QM2}} (0)
\end{array}}}_{\equiv \mb{C}(0)},
}
where $ \tilde s \equiv s/\gamma$ and the matrix $\dbar M (\tilde s)$  is defined as:
\eqn{&\dbar M (\tilde s) = \non\\
&\bkt{\begin{array}{ccccccc}
     \tilde s + \frac{1}{2}& 0&0&\frac{\sqrt{N}}{2} e^{i k_0 (x_1 + x_A)} &0& \frac{\sqrt{N}}{2} e^{i k_0 (x_1 - x_A)}&0\\
     0 & \tilde s + \frac{1}{2} & 0 & 0& \frac{\sqrt{N}\gamma}{2} e^{i k_0 (x_1 + x_A)}& 0 &0\\
          0& 0&\tilde s + \frac{1}{2}&0 &0& 0&\frac{\sqrt{N}}{2} e^{i k_0 (x_1 - x_A)}\\  \frac{\sqrt{N}}{2} e^{i k_A (x_1 + x_A)}& 0&0&\tilde s + \frac{N}{2}&0 &\frac{N}{2}e^{2 i k_0 x_1}& 0\\
          0& \frac{\sqrt{N}}{2} e^{i k_A (x_1 + x_A)}&0&0 &\tilde s + \frac{N}{2}& 0\\
          \frac{\sqrt{N}}{2} e^{i k_A (x_1 - x_A)}& 0&0&\frac{N}{2}e^{2 i k_0 x_1} &0&\tilde s + \frac{N}{2}& 0\\
          0& 0&\frac{\sqrt{N}}{2} e^{i k_A (x_1 - x_A)} &0& 0&0&\tilde s + \frac{N}{2}
\end{array}}
}

The solution to the atomic dynamics can be thus numerically obtained in terms of the inverse Laplace $ \mb{C}(t) = \mc{L}^{-1} \sbkt{\frac{1}{\gamma}\dbar M^{-1}(\tilde s) \mb{C}(0)}$.

\subsection{Large $N$ limit}

Let us consider the dynamics in the subspace where both QMs are in the ground state $ \ket{G}$ (Eqs.\eqref{eq:cagg}, \eqref{eq:ceg}, \eqref{eq:cge}). Using the initial conditions $ c_\mr{QM1} (0) = c_\mr{QM2} (0) = 0 $ in Eq.\eqref{eq:ceg} and  \eqref{eq:cge}, yields:
\eqn{\label{eq:cegs}
\tilde c_\mr{QM1}(s) = - \frac{\sqrt{N}\gamma}{2}e^{i k_0 x_1 } \frac{\sbkt{e^{i k_0 x_A } \bkt{s + \frac{N \gamma}{2}} - e^{- i k_0 (x_A - 2 x_1) } \frac{N \gamma}{2} }}{\bkt{s + \frac{N \gamma}{2}}^2 - \bkt{\frac{N \gamma}{2}e^{2 i k_0 x_1}}^2}\tilde c_{\mr{A},GG}  (s) \\
\label{eq:cges}
\tilde c_\mr{QM2}(s) = - \frac{\sqrt{N}\gamma}{2}e^{i k_0 x_1 } \frac{\sbkt{e^{-i k_0 x_A } \bkt{s + \frac{N \gamma}{2}} - e^{ i k_0 (x_A + 2 x_1) } \frac{N \gamma}{2} }}{\bkt{s + \frac{N \gamma}{2}}^2 - \bkt{\frac{N \gamma}{2}e^{2 i k_0 x_1}}^2}\tilde c_{\mr{A},GG} (s) 
}
Substituting the above in Eq.\eqref{eq:cagg}, we obtain:
\eqn{\label{eq:tildeca}
\tilde c_{\mr{A},GG} (s)  = c_{\mr{A},GG} (0)\sbkt{\frac{\bkt{s + \frac{N\gamma}{2}}^2 - \bkt{\frac{N \gamma}{2} e^{2 i k_0 x_1}}^2}{\bkt{s + \frac{\gamma}{2}} \cbkt{\bkt{s + \frac{N\gamma}{2}}^2 - \bkt{\frac{N \gamma}{2} e^{2 i k_0 x_1}}^2} -  \frac{N \gamma^2 }{2} e^{2 i k_0 x_1} \cbkt{ \cos \bkt{2 k_0 x_A} \bkt{s + \frac{ N \gamma}{2}} - \frac{N \gamma}{2}e^{2 i k_0 x_1}}}}.
}

We now consider the large $N$ limit for the QMs for following cases as discussed in the main text:
\begin{enumerate}
    \item {Atom at cavity node $(x_A = 0, x_1 = p \lambda_0 /2, p\in \mathbb{N})$: One can simplify Eq.\eqref{eq:tildeca} in this case as follows 
    \eqn{\label{eq:cagg1}
    \tilde c_{\mr{A},GG } (s) = c_{\mr{A},GG} (0)  \sbkt{\frac{{s + N\gamma}}{\bkt{s + \frac{\gamma}{2}} \bkt{s + N \gamma }  -  \frac{N \gamma^2 }{2}  }},
    }
    such that in the large $N$ limit this can be approximated as:
    \eqn{
        \tilde c_{\mr{A},GG } (s) \approx \frac{1}{s}c_{\mr{A},GG} (0) \implies c_{\mr{A},GG }(t)  = c_{\mr{A},GG} (0). 
    }
    Substituting  Eq.\eqref{eq:cagg1} in Eq.\eqref{eq:cegs} and \eqref{eq:cges}, we get $ \tilde c_\mr{QM1}(s) = \tilde c_\mr{QM2}(s) = - \frac{\sqrt{N}\gamma}{2} \frac{1}{s + N \gamma}\tilde c_{\mr{A},GG} (s) $, such that in the large $N$ limit both  $c_\mr{QM1}(t), c_\mr{QM2}(t)\propto 1/\sqrt{N}\rightarrow 0 $. 

    }
    \item {Atom at cavity antinode $(x_A = 0, x_1 = p \lambda_0 /4,  p\in \mathbb{N})$: In this case Eq.\eqref{eq:tildeca} becomes:
    \eqn{\label{eq:cagg2}
\tilde c_{\mr{A},GG} (s)  = c_{\mr{A},GG} (0) \frac{s}{s^2 + \frac{\gamma}{2} s +  \frac{N \gamma^2 }{2} },
}
which can be approximated in the large $N $ limit as:
    \eqn{
& \tilde c_{\mr{A},GG} (s)  \approx \frac{c_{\mr{A},GG} (0)}{2}\sbkt{\frac{1}{s+ \frac{\gamma}{4} + i \sqrt{\frac{N}{2}} \gamma}  + \frac{1}{s + \frac{\gamma}{4} - i \sqrt{\frac{N}{2}} \gamma} } \\
\implies & c_{\mr{A }, GG} (t)  \approx c_{\mr{A},GG} (0) e^{- \gamma t/4}\cos \bkt{\sqrt{\frac{N}{2}}\gamma t}.
}
Substituting Eq.\eqref{eq:cagg2} in Eq.\eqref{eq:cegs} and \eqref{eq:cges} yields:
\eqn{
&\tilde c_\mr{QM1}(s) = \tilde c_\mr{QM2}(s) = - \frac{i\sqrt{N}\gamma}{2\bkt{s^2 + \frac{\gamma}{2}s + \frac{N \gamma ^2 }{2}}} c_{\mr{A},GG}(0) 
}
Taking the large $N$ limit, 
\eqn{&\tilde c_\mr{QM1}(s) =  \tilde c_\mr{QM2}(s)  \approx \frac{c_{\mr{A},GG}(0) }{\sqrt{2}} \sbkt{\frac{1}{s + \frac{\gamma}{4}+ i \sqrt{\frac{N}{2}} \gamma}  - \frac{1}{s+ \frac{\gamma}{4} - i \sqrt{\frac{N}{2}} \gamma }}\\
\implies& c_\mr{QM1}(t) = c_\mr{QM2}(t) \approx \sqrt{2}i c_{\mr{A},GG}(0) e^{- \gamma t /4} \sin \bkt{\sqrt{\frac{N}{2}}\gamma t}}
}
    \item {Atom near cavity node $(0<k_0 x_A \ll 1 , x_1 = p \lambda_0 /2,  p\in \mathbb{N})$: We define $\phi_A \equiv k_0 x_A $, such that $ \cos \bkt{2k_0 x_A }\approx 1 - 2\phi_A ^2 $, thus Eq.\eqref{eq:tildeca} becomes:

    \eqn{ \label{eq:cagg3}\tilde c_{\mr{A},GG} (s)  
    \approx&c_{\mr{A},GG} (0)\sbkt{\frac{s\bkt{s + N\gamma}}{s\bkt{s + \frac{\gamma}{2}} \bkt{s + N\gamma}-  \frac{N \gamma^2 }{2} \cbkt{ s  - 2 \phi_A ^2\bkt{s + \frac{ N \gamma}{2}}}}}
    }
    Taking the large $N$ limit:
    \eqn{ \tilde c_{\mr{A},GG} (s)  
    \approx&c_{\mr{A},GG} (0)\sbkt{\frac{s }{s^2  +  \phi_A ^2\frac{ N \gamma^2}{2}}}\implies c_{\mr{A},GG }(t) \approx c_{\mr{A},GG }(0) \cos \bkt{\phi_A \sqrt{\frac{N}{2}}\gamma t}
    }
    To obtain the dynamics of the  atoms constituting the QMs, we substitute Eq.\eqref{eq:cagg3} in Eqs.\eqref{eq:cegs} and \eqref{eq:cges}:
    \eqn{
\tilde c_\mr{QM1}(s) \approx - \frac{\sqrt{N}\gamma}{2}  \frac{e^{i k_0 x_A } \bkt{s + \frac{N \gamma}{2}} - e^{- i k_0 x_A }\frac{N \gamma}{2} }{s\bkt{s + \frac{\gamma}{2}} \bkt{s + N\gamma}-  \frac{N \gamma^2 }{2} \cbkt{ s  - 2 \phi_A ^2\bkt{s + \frac{ N \gamma}{2}}}} c_{\mr{A},GG}(0) \\
\tilde c_\mr{QM2}(s) \approx- \frac{\sqrt{N}\gamma}{2}\frac{e^{-i k_0 x_A } \bkt{s + \frac{N \gamma}{2}} - e^{i k_0 x_A } \frac{N \gamma}{2} }{s\bkt{s + \frac{\gamma}{2}} \bkt{s + N\gamma}-  \frac{N \gamma^2 }{2} \cbkt{ s  - 2 \phi_A ^2\bkt{s + \frac{ N \gamma}{2}}}} c_{\mr{A},GG}(0)
}
Taking the large $N$ limit of the above, we get:
\eqn{
&\tilde c_\mr{QM1}(s) \approx - \tilde c_\mr{QM2}(s)\approx- \frac{\sqrt{N}\gamma}{2}  \frac{i \phi _A}{s^2+  \frac{N \gamma^2 }{2}   \phi_A ^2  } c_{\mr{A},GG}(0) \\
\implies&  c_\mr{QM1}(t) \approx -  c_\mr{QM2}(t)\approx-  \frac{i }{\sqrt{2}} \sin  \bkt{\phi_A\sqrt{\frac{N}{2}}\gamma t}
    .}}
\end{enumerate}

\end{document}